\documentclass[preprint,showpacs,amssymb,aps,superscriptaddress,pre]{revtex4}
\usepackage{graphicx}
\usepackage{dcolumn}
\usepackage{bm}

\begin{document}
\title{Phase diagrams of the spin-1/2 Ising-Heisenberg model on a triangle-hexagon lattice}
\author{J. KI\v{S}\v{S}OV\'{A}} 
\author{J. STRE\v{C}KA} 
\affiliation{Department of Theoretical Physics and Astrophysics, Faculty of Science, \\ 
P. J. \v{S}af\'{a}rik University, Park Angelinum 9, 040 01 Ko\v{s}ice, Slovak Republic} 
      
\begin{abstract}
Ground-state and finite-temperature phase diagrams of a geometrically frustrated spin-1/2 Ising-Heisenberg model on a triangle-hexagon lattice are investigated within the generalized star-triangle mapping transformation. It is shown that the ground state is constituted by two different spontaneously long-range ordered phases and one disordered spin-liquid phase.
\end{abstract}
\pacs{05,30,Rt; 05.50.+q; 05.70.Jk; 75.10.Kt; 75.10.Jm; 75.30.Kz} 

\maketitle

\section{Introduction}
The spin-1/2 Ising-Heisenberg model on geometrically frustrated planar lattices attracts a great deal of research interest as it may exhibit both spontaneous long-range order as well as peculiar spin-liquid phases \cite{stre06}. In this work, the ground-state and finite-temperature phase diagrams of the spin-1/2 Ising-Heisenberg model on a triangle-hexagon lattice will be investigated in detail.

\section{Model and its exact solution}
Consider the spin-1/2 Ising-Heisenberg model on the triangle-hexagon lattice which is schematically illustrated in Fig.~\ref{fig1}. In this figure, the full circles denote lattice positions of the Ising spins $\sigma=1/2$, while the empty ones label lattice positions of the Heisenberg spins $S=1/2$. For further convenience, the total Hamiltonian can be defined as a sum of the cluster Hamiltonians, i.e. 
$\hat{\cal H}= \sum_{k} {\hat{\cal H}_k}$, where each cluster Hamiltonian $\hat{\cal H}_k$ involves all the interaction terms of the Heisenberg trimer from the $k$-th triangle unit cell depicted in Fig. 1
\begin{eqnarray}
\hat{\cal H}_{k} = \! \! \! &-& \! \! \! J_{\rm H} \sum_{i=1}^3 [\Delta (\hat{S}_{k,i}^x \hat{S}_{k,i+1}^x  + \hat{S}_{k,i}^y\hat{S}_{k,i+1}^y) + \hat{S}_{k,i}^z \hat{S}_{k,i+1}^z] \nonumber \\
\! \! \! &-& \! \! \! J_{\rm I} \sum_{i=1}^3 \hat{S}_{k,i}^z \hat{\sigma}_{k,i}^z.
\label{fig1}
\end{eqnarray}
In above, $\hat{S}_{ki}^{\alpha}$ and $\hat{\sigma}_{ki}^{\alpha}$ ($\alpha=x,y,z$; $i=1,2,3$) label spatial components of the spin-1/2 operator, the parameter $J_{\rm I}$ represents the Ising-type interaction between the nearest-neighbour Ising and Heisenberg spins, the parameter $J_{\rm H}$ labels the Heisenberg-type interaction between the nearest-neighbour Heisenberg spins in the inner triangles, $\Delta$ is a spatial anisotropy in this latter interaction and $\hat{S}_{k,4}^{\alpha} = \hat{S}_{k,1}^{\alpha}$ is implied for convenience.
\begin{figure}
\vspace{-0.3cm}
\includegraphics[width=9.5cm]{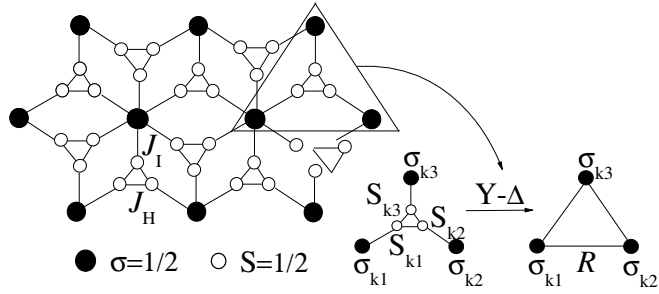}
\vspace{-1.7cm}
\caption{A segment of the spin-1/2 Ising-Heisenberg model on the triangle-hexagon lattice. The figure on the right schematically shows the star-triangle transformation.}
\label{fig1}
\end{figure}

A crucial step of our approach lies in evaluation of the partition function. Due to a commutability of different bond Hamiltonians, the partition function can be partially factorized to a product
\begin{eqnarray}
{\cal Z}_{\rm IHM}=\sum_{\{ \sigma_i \}} \prod_{k=1}^{2N} {{\rm Tr}_k \exp (-\beta \hat{\cal H}_k)} = \sum_{\{ \sigma_i \}} \prod_{k=1}^{2N} {\cal Z}_k,
\label{2}
\end{eqnarray}
where $\beta = 1/(k_{\rm B} T)$, $k_{\rm B}$ is Boltzmann's constant, $T$ denotes the absolute temperature and $N$ is the total number of the Ising spins. Furthermore, the symbol ${\rm Tr}_k$ stands for a trace over degrees of freedom of the $k$-th Heisenberg trimer and the indicated summation runs over all available configurations of the Ising spins. 

To find an explicit form of the partition function ${\cal Z}_k$, it is sufficient to consider only two symmetry-distinct configurations of the Ising spins enclosing the $k$-th Heisenberg trimer and to diagonalize the relevant Hamiltonian (\ref{1}) for both independent spin configurations. In this way, one obtains two different cluster partition functions, the one ${\cal Z}_{k} (1/2,1/2,1/2)$ for all three Ising spins aligned parallel and the one ${\cal Z}_{k} (1/2,1/2,-1/2)$ for one of three Ising spins aligned antiparallel with respect to the other two. These expressions are rather cumbersome and hence, they are not explicitly given here for brevity.

Subsequently, the partition function ${\cal Z}_k$ can be substituted by means of the generalized star-triangle transformation \cite{roja09}, as schematically depicted in Fig.~\ref{fig1} 
\begin{eqnarray}
{\cal Z}_k (\sigma_{k1}^z, \sigma_{k2}^z, \sigma_{k3}^z) \! = \! A \exp \left( \beta R \sum_{i=1}^{3} \sigma_{k,i}^z \sigma_{k,i+1}^z \right),
\label{3}
\end{eqnarray}
where $\sigma_{k,4}^{z}=\sigma_{k,1}^{z}$ is implied here for convenience.
Considering only two symmetry-distinct configurations of three Ising spins enclosing the $k$-th Heisenberg trimer, the mapping parameters $A$ and $R$ can be expressed with the help of both independent cluster partition functions
\begin{eqnarray}
A \! \! \! &=& \! \! \! [{\cal Z}_k (1/2,1/2,1/2) \, {\cal Z}_k (1/2,1/2,-1/2)]^{1/4} \\ 
2 \beta R \! \! \! &=& \! \! \! \ln [{\cal Z}_k (1/2,1/2,1/2)] - \ln [{\cal Z}_k (1/2,1/2,-1/2)].
\label{4}
\end{eqnarray}
The precise mapping relationship between the partition function of the spin-1/2 Ising-Heisenberg model on the triangle-hexagon lattice and the partition function of a simple spin-1/2 Ising model on the triangular lattice can be obtained by direct substitution of the star-triangle transformation (\ref{3}) into the formula (\ref{2}) 
\begin{eqnarray}
{\cal Z}_{\rm IHM} (\beta, J_{\rm I}, J_{\rm H}, \Delta) = A^{2N} {\cal Z}_{\rm IM} (\beta, R).
\label{5}
\end{eqnarray}
Our exact calculation of the partition function ${\cal Z}_{\rm IHM}$ is thus formally completed, since the exact result for the partition function of the spin-1/2 Ising model on a triangular lattice is well known \cite{hout50}. Note that all the other thermodynamic quantities can also be calculated in a rather straightforward way using the basic relations of thermodynamics and statistical physics.

\section{Results and discussion}

First, let us focus on the ground state of the spin-1/2 Ising-Heisenberg model on the triangle-hexagon lattice. The ground state of this geometrically frustrated spin system consists of two different spontaneously long-range ordered phases and one disordered spin-liquid phase. The disordered spin-liquid phase with a non-zero residual entropy appears whenever antiferromagnetic interaction between the Heisenberg spins ($J_{\rm H}<0$) is assumed, because the outer Ising spins become geometrically frustrated on behalf of the frustrated character of the Heisenberg spins from the inner triangles. For the ferromagnetic Heisenberg interaction ($J_{\rm H}>0$), the Ising spins always exhibit a perfect ferromagnetic long-range order, while the Heisenberg spins are ferromagnetically aligned with respect to each other in the classical ferromagnetic phase (CFP) or display a remarkable two-up one-down spin arrangement in the quantum ferromagnetic phase (QFP). The phase boundary between CFP and QFP is given by the constraint
\begin{eqnarray}
\Delta_{\rm b} = 1 + |J_{\rm I}|/2 J_{\rm H}.
\end{eqnarray}
The CFP occurs whenever $\Delta \leq \Delta_{\rm b}$, while the QFP appears for any $\Delta \geq \Delta_{\rm b}$. 

\begin{figure}[thb]
\vspace{1.0cm}
\includegraphics[width=6cm]{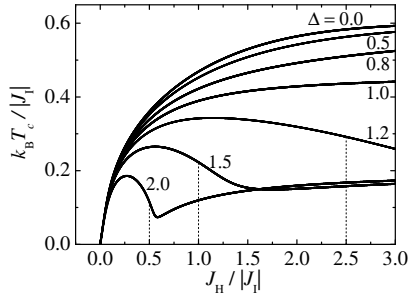}
\vspace{-0.7cm}
\caption{Critical temperature as a function of $J_{\rm H}/|J_{\rm I}|$ for several values of the exchange anisotropy $\Delta$.}
\label{fig2}
\end{figure}

Next, let us discuss the dependence of the critical temperature on the ratio $J_{\rm H}/|J_{\rm I}|$ between the Heisenberg and Ising interactions. The critical temperature can easily be obtained by comparing the effective interaction $\beta R$ with the respective critical value $\beta_{\rm c} R = \ln 3$ of the corresponding spin-1/2 Ising model on a triangular lattice \cite{hout50}. As one can see from Fig.~\ref{fig2}, the critical temperature is a monotonically decreasing function of the ratio $J_{\rm H}/|J_{\rm I}|$ for the easy-axis exchange anisotropies ($\Delta \leq 1$), while it becomes a more interesting non-monotonic dependence that reflects a crossover from CFP to QFP whenever the easy-plane exchange anisotropies ($\Delta>1$) are considered. 
The dependences of critical temperature converge to three different limiting values in the limit of $J_{\rm H}/|J_{\rm I}| \to \infty$ for $\Delta>1$, $\Delta<1$ and in special case of $\Delta=1$. For $\Delta<1$, the critical temperature tends to the value $\frac{k_{\rm B} T_{\rm c}}{|J_{\rm I}|} = \frac{1}{2\ln[(1+\sqrt{3}+\sqrt{2}\sqrt[4]{3})/2]}$, for $\Delta>1$ the critical temperature converges to the value $\frac{k_{\rm B} T_{\rm c}}{|J_{\rm I}|} = \frac{1}{6\ln[(1+\sqrt{3}+\sqrt{2}\sqrt[4]{3})/2]}$. The straight vertical lines in the Fig.~\ref{fig2} represent the phase boundaries between CFP and QFP for three different values of the easy-plane exchange anisotropy $\Delta>1$.

In conclusion, the exact solution for the spin-1/2 Ising-Heisenberg model on the triangle-hexagon lattice has been derived within the framework of the generalized star-triangle transformation, which has been used for a detailed analysis of the ground-state and finite-temperature phase diagrams. 
In the future, we plan to complete the present study by calculating other thermodynamic quantities
confirming aforedescribed scenario.

{\bf Acknowledgments}: The financial support provided under the grants VVGS 2/09-10 and VEGA 1/0431/10 is gratefully acknowledged.

\end{document}